\documentclass[conference]{IEEEtran}

\usepackage{graphics,graphicx}
\usepackage{url} 						
\usepackage{times,tikz}
\usepackage{color}
\usepackage{xcolor}
\usepackage{tabulary}
\usepackage[utf8x]{inputenc}
\usepackage[cmex10]{amsmath}
\usepackage{cite, amsfonts, amssymb, amsthm, bm, bbm, graphicx,  multirow,  tikz,subfigure}
\usepackage[american]{babel}
\usepackage{algorithmic, algorithm}
\usepackage[multiple]{footmisc}

\usepackage[numbers]{natbib}

\newcommand{\ds}{\displaystyle}

\usepackage{graphicx}

\ifCLASSINFOpdf
\else
\fi

\begin{document}
\title{Doppler-Resilient Universal Filtered MultiCarrier (DR-UFMC): A Beyond-OTFS Modulation}

\author{\IEEEauthorblockN{Carmen D'Andrea}
\IEEEauthorblockA{\textit{University of Cassino}\\
\textit{and Southern Latium}\\
Cassino, Italy\\
carmen.dandrea@unicas.it}
\and
\IEEEauthorblockN{Stefano Buzzi}
\IEEEauthorblockA{\textit{Univ. of Cassino and Southern} \\
	\textit{Latium}, Cassino, Italy, and  \\
	\textit{Politecnico di Milano}, Milan, Italy \\
	buzzi@unicas.it}
\and
\IEEEauthorblockN{Maria Fresia, Xiaofeng Wu}
\IEEEauthorblockA{\textit{Huawei Technologies Duesseldorf GmbH} \\ \textit{Wireless Terminal Chipset Technology Lab} \\ Munich, Germany \\
\{maria.fresia, xiaofeng.wu\}@huawei.com}}

\maketitle

\begin{abstract}
In the past few years, some alternatives to the Orthogonal Frequency Division Multiplexing (OFDM) modulation have been considered to improve its spectral containment and its performance level in the presence of heavy Doppler shifts. This paper examines a novel modulation, named Doppler-Resilient Universal Filtered MultiCarrier (DR-UFMC), which has the objective of combining the advantages provided by the Universal Filtered MultiCarrier (UFMC) modulation (i.e., better spectral containment), with those of the Orthogonal Time Frequency Space (OTFS) modulation (i.e., better performance in time-varying environments). The paper contains the mathematical model and detailed transceiver block scheme of the newly described modulation, along with a numerical analysis contrasting DR-UFMC against OTFS, OFDM with one-tap frequency domain equalization (FDE), and OFDM with multicarrier multisymbol linear MMSE processing. Results clearly show the superiority, with respect to the cited benchmarks,  of the newly proposed modulation in terms of achievable spectral efficiency. Interestingly, it is also seen that OFDM, when considered in conjunction with multicarrier multisymbol linear minimum mean squares error (MMSE) processing, performs slightly better than OTFS in terms of achievable spectral efficiency.
\end{abstract}


\IEEEpeerreviewmaketitle

\section{Introduction}
Future 6G wireless networks are being envisioned with several challenging use cases, including communication in environments with extreme mobility. Examples of such scenarios include high-speed trains, drones, and communications through non-terrestrial networks. The fast movement of at least one communication terminal creates rapid variations in the channel and reduces the coherence time, which can negatively impact the performance of traditional OFDM modulation with one-tap FDE. In order to overcome such limitation, the OTFS modulation has been proposed few years ago  by researchers of the Silicon Valley tech company Cohere Technologies \cite{hadani2017orthogonal,hadani2018otfs}. OTFS is a two-dimensional modulation scheme in which information symbols are multiplexed in the delay–Doppler domain; it is claimed to be robust against both frequency and time selectivity of the wireless mobile channel since each information symbol is spread, by means of a 2-D Fourier transform, on a grid of the time-frequency plane. 
As recognized in \cite{Heath2019OTFS}, OTFS modulation can be realized by adding a pre-processing block before a traditional modulator in the frequency-time domain such as the OFDM modulator at the transmitter, and a corresponding post-processing block after a traditional demodulator in the frequency-time domain such as OFDM demodulator at the receiver. In its basic formulation, thus, OTFS still retains many of the disadvantages of OFDM in terms of out-of-band (OOB) emission and, in some implementations, need for a cyclic prefix (CP). Reference \cite{gaudio2021otfs} is one of the first research work providing a fair comparison between OTFS and its direct competitor and widely used OFDM. The authors present such a fair comparison between the two digital modulation formats in terms of achievable communication rate also addressing the problem of channel estimation for the two modulations.  
In \cite{raviteja2018interference}, authors  derive the explicit input-output relation describing OTFS modulation and demodulation (mod/demod). Then the cases of ideal pulse-shaping waveforms (that satisfy the bi-orthogonality conditions) and rectangular waveforms (which do not) are analyzed. The study also shows through numerical results the superior error performance gains of the proposed uncoded OTFS schemes over OFDM under various channel conditions.

As already commented, while OTFS generally outperforms OFDM in the presence of heavy Doppler shifts, in its basic implementation suggested in \cite{Heath2019OTFS} it shares with OFDM a reduced spectral efficiency. 
In this paper, to circumvent this drawbacks, we explore a novel modulation format, that we name Doppler-Resilient Universal Filtered MultiCarrier (DR-UFMC) modulation, that merges together the advantages of OTFS (i.e., robustness in time-varying environments) with those of the UFMC modulation \cite{knopp2016universal,buzzi2019mimo}, i.e. better spectral containment and no need for the CP. Specifically, this paper makes the following contribution: first of all the full continuous-time mathematical derivation of the transmitted and received signal for the DR-UFMC modulation is developed; next, a full comparison of the DR-UFMC modulation versus OTFS, OFDM with one-tap FDE and with multicarrier multisymbol processing is presented for time-varying environment, in terms of achieved Signal-to-Interference plus Noise-Ratio (SINR), normalized Mean Square Error (MSE), and Achievable Spectral Efficiency under OOB emission constraints.  

This paper is organized as follows. Next section contains a review of the OTFS modulation scheme, while Section III is devoted to the exposition of the newly proposed DR-UFMC modulation. Section IV contains the derivation of the expressions for the considered performance measures, and presents the numerical results of the paper. Finally, concluding remarks are given in Section V. 

\begin{figure*}[t]
	\begin{center}
		\includegraphics[scale=0.28]{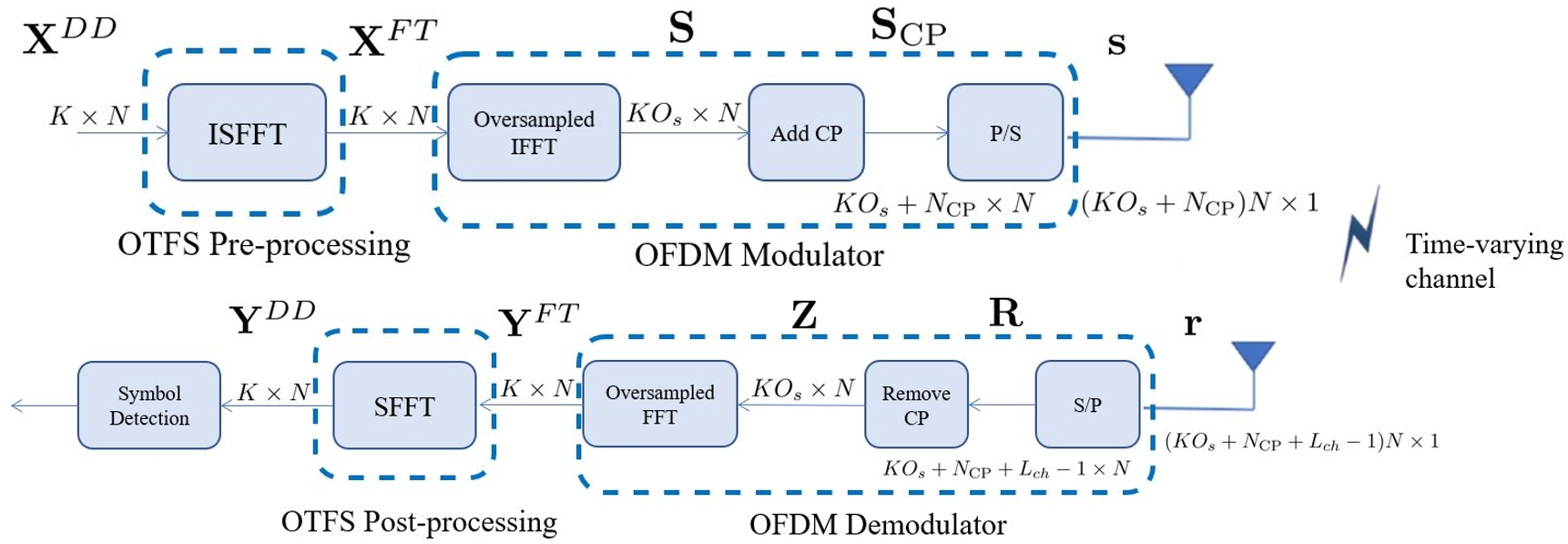}
		\caption{OTFS transceiver architecture with oversampled FFT and IFFT operations}
	\end{center}
	\label{Fig:OTFS_block_scheme}
\end{figure*}

\section{A brief review of OTFS}

Fig. 1 shows a block scheme of the OTFS transceiver.
The matrix $\mathbf{X}^{DD}$ of size $K \times N$ contains the $KN$ information symbols in the {delay-Doppler domain},  where $K$ and $N$ are the numbers of resource units along the delay dimension and Doppler dimension, respectively.
The {pre-processing block} performs an {inverse symplectic finite Fourier transform} (ISFFT) as follows:
\begin{equation}
	\mathbf{X}^{FT}= \mathbf{F}_K \mathbf{X}^{DD} \mathbf{F}_N^H
	\label{X_ISFFT}
\end{equation}
where $\mathbf{F}_K$ and $\mathbf{F}_N$ represent the isometric DFT matrices of size $K$ and $N$, respectively. 
Then, the 2D block $\mathbf{X}^{FT}$ in the frequency-time domain is transformed to the 1-D transmit signal through a traditional OFDM modulator, with an oversampled IFFT operation to emulate continuous time subcarriers. Denoting by $O_s$ the oversampling factor, we have
\begin{equation}
	\mathbf{S}=\overline{\mathbf{W}}_{K,O_s}^H \mathbf{X}^{FT}\, ,
	\label{S_OFDM1} 
\end{equation}
where the $(\ell, m)$-th entry of the matrix $\overline{\mathbf{W}}_{K,O_s}$ is defined as
\begin{equation}
\overline{\mathbf{W}}_{K,O_s}(\ell,m)=\ds \frac{1}{\sqrt{KO_s}}e^{-j2\pi\frac{(m-1)(\ell-K/2-1)}{KO_s}}
\end{equation}
with $\ell=1,\ldots, K$ and $m=1,\ldots, KO_s$.
Each column of the matrix $\mathbf{S}$ can be thus seen as an OFDM symbol.
Combining Eqs. \eqref{X_ISFFT}-\eqref{S_OFDM1}, we have:
$	\mathbf{S}=\overline{\mathbf{W}}_{K,O_s}^H \mathbf{F}_K \mathbf{X}^{DD} \mathbf{F}_N^H$.
The CP is included in the transmitted signal through the matrix 
\begin{equation}
	\bar{\mathbf{A}}_{N_{\rm CP}} = [\mathbf{I}_{KO_s}(KO_s-N_{CP}+1:KO_s,:) ; \mathbf{I}_{KO_s}],
	\label{A_CP_definition_oversampled}
\end{equation}
with $\mathbf{I}_x$ the identity matrix of order $x$.  
Letting $\mathbf{S}_{\rm CP}=	\bar{\mathbf{A}}_{N_{\rm CP}} \mathbf{S}$, the 1-D signal to be transmitted is obtained through columnwise vectorization of the matrix $\mathbf{S}_{\rm CP}$, i.e.: 
$\mathbf{s} = \text{vec}\{\mathbf{S}_{\rm CP}\}$.
At the receiver, the $(KO_s+N_{\rm CP}+L_{ch}-1) N$-th dimensional received signal $\mathbf{r}$ can be written as
\begin{equation}
	\mathbf{r}= \sqrt{P_T} \overline{\mathbf{M}}_{h,N}\mathbf{s}+\mathbf{w},
	\label{eq:received_OTFS}
\end{equation}
where $P_T$ is the transmitted power, $\mathbf{w}$ is the AWGN contribution, while $\overline{\mathbf{M}}_{h,N}$ describes the channel effect. In particular, denoting by 
$g(\cdot)$ the continuous-time convolution of the transmit and receive shaping filters, by $\nu_p$ the Doppler shift associated to the $p$-th path of the channel, by $h_p$  the amplitude of the $p$-th path of the channnel, and letting $\widetilde{T}_s=T_s/O_s$, with $T_s$ the OFDM signaling interval $1/(K \Delta_f)$, the time-variant channel can be expressed as \cite{Heath2019OTFS} 
\begin{equation}
	h^{(i)}_{r,\ell}= \ds \sum_{p=0}^{P−1} h_p g(\ell \widetilde{T}_s-\tau_p) e^{j2\pi \nu_p [(\ell + r +i-1)\widetilde{T}_s-\widetilde{T}_s/2]},
	\label{LTV_channel_i_exp}
\end{equation}
for $\ell=1,\ldots, L_{ch}$, $i=1,\ldots, N$. Based on the above notation, the channel matrix in \eqref{eq:received_OTFS} is expressed as  
$\overline{\mathbf{M}}_{h,N}=\text{blckdiag}\left(\mathbf{M}_{h}^{(1)},\ldots, \mathbf{M}_{h}^{(N)} \right)$, where the $(r,c)$-th entry of  the  $\left[(KO_s+L_{ch}-1) \times (KO_s + N_{\rm CP})\right]$-dimensional matrix $\mathbf{M}_h^{(i)}$ are defined as 
\begin{equation}
	\mathbf{M}_h^{(i)}(r,c)=\left \lbrace
	\begin{array}{llll}
		h^{(i)}_{r,r-c} & \text{if} \; 0<r-c \leq L_{ch}
		\\
		0 & \text{otherwise} 
	\end{array}\right.
	\label{M_h_i_LTV_channel}
\end{equation}

The OTFS demodulation at the receiver consists of the traditional OFDM demodulator and a {post-processing block}. The OFDM demodulator transforms the received signal $\mathbf{r}$ into a 2-D block in the frequency-time domain $\mathbf{Y}^{FT}$ with dimensions $K\times N$. Specifically, the vector $\mathbf{r}$ is first rearranged as a matrix $\mathbf{R} $ of size $(KO_s+N_{\rm CP} +L_{ch}-1 )\times N$, i.e.,
$\mathbf{R} = \text{invec}\{\mathbf{r}\}$,
where each column vector of $\mathbf{R}$ can be interpreted as a received OFDM symbol. Then, the CP is removed through the CP removal matrix $	\overline{\mathbf{R}}_{N_{\rm CP}}$, defined as
$
	\overline{\mathbf{R}}_{N_{\rm CP}}=\left[ \mathbf{0}_{KO_s\times N_{\rm CP}} \; , \mathbf{I}_{KO_s \times KO_s+ L_{ch}-1}
	\right]
$,
and the received signal after CP removal can be written as $\mathbf{Z}=\overline{\mathbf{R}}_{N_{\rm CP}}\mathbf{R}$.
It can be shown that the $i$-th column of $\mathbf{Z}$ is expressed as
\begin{equation}
	\begin{array}{llll}
	\mathbf{Z}(:,i)= &\sqrt{P_T} \overline{\mathbf{H}}_{i} \mathbf{S}(:,i) + \mathbf{V}(:,i) = \\ & \sqrt{P_T} \overline{\mathbf{H}}_{i} \overline{\mathbf{W}}_{K,O_s}^H \mathbf{F}_K \mathbf{X}^{DD} \mathbf{F}_N^*(:,i) + \mathbf{V}(:,i)
	\end{array}
	\label{Z_i}
\end{equation}
where $\mathbf{V}=\overline{\mathbf{R}}_{N_{\rm CP}}\mathbf{W}$ and 
$\overline{\mathbf{H}}_{i}=\overline{\mathbf{R}}_{N_{\rm CP}}\mathbf{M}_h^{(i)} \bar{\mathbf{A}}_{N_{\rm CP}}$ is a  $(KO_s\times KO_s)$-dimensional matrix. 
Applying the $KO_s$-point oversampled FFT on each OFDM symbol, i.e., on each column of the matrix $\mathbf{Z}$, we obtain the received 2-D $K\times N$-dimensional block  $
	\mathbf{Y}^{FT} = \overline{\mathbf{W}}_{K,O_s} \mathbf{Z}$ in the frequency-time domain. Then, in the post-processing block, $\mathbf{Y}^{FT}$ is transformed into the 2-D data block $\mathbf{Y}^{DD}$ in the delay-Doppler domain through  a {symplectic finite Fourier transform} (SFFT): 
\begin{equation}
	\begin{array}{lll}
	\mathbf{Y}^{DD}= & \mathbf{F}_K^H \mathbf{Y}^{FT} \mathbf{F}_N= 
	\mathbf{F}_K^H \overline{\mathbf{W}}_{K,O_s}  \mathbf{Z}\mathbf{F}_N= \\ &
	\mathbf{F}_K^H \overline{\mathbf{W}}_{K,O_s}\ds \sum_{i=1}^N \mathbf{Z}(:,i) \mathbf{F}_N(:,i)^T
\end{array}
	\label{Y_DD}
\end{equation}
By substituting \eqref{Z_i} into \eqref{Y_DD} we obtain
\begin{equation} \begin{array}{lll}
	\mathbf{Y}^{DD}=  &
	\sqrt{P_T} \ds \sum_{i=1}^N \mathbf{F}_K^H \overline{\mathbf{W}}_{K,O_s} \overline{\mathbf{H}}_{i} \overline{\mathbf{W}}_{K,O_s}^H \mathbf{F}_K \mathbf{X}^{DD} \, \times \\ & \mathbf{F}_N^*(:,i) \mathbf{F}_N(:,i)^T  + \mathbf{V}_{DD}
\end{array} \label{Y_DD2}
\end{equation}
with $\mathbf{V}_{DD}=\mathbf{F}_K^H \overline{\mathbf{W}}_{K,O_s} \mathbf{V}\mathbf{F}_N$.
We now denote the $(k,n)$-th element of $\mathbf{Y}^{DD}$, $\mathbf{X}^{DD}$ and $\mathbf{V}^{DD}$ as $Y_{k,n}^{DD}$, $X_{k,n}^{DD}$, $V_{k,n}^{DD}$, respectively, where $k=0,\ldots K-1$ and $n= 0 ,\ldots, N-1$.
Upon defining $	\overline{\mathbf{B}}_{i}=\mathbf{F}_K^H \overline{\mathbf{W}}_{K,O_s} \overline{\mathbf{H}}_{i} \overline{\mathbf{W}}_{K,O_s}^H \mathbf{F}_K$,
the following expression can be shown to be obtained for $Y_{k,n}^{DD}$:
\begin{equation}
	\begin{array}{lll}
	Y_{k,n}^{DD}=\\ \ds \frac{\sqrt{P_T}}{N}\ds \sum_{k'=0}^{K-1} \ds \sum_{n'=0}^{N-1} X_{k',n'}^{DD} \ds \sum_{i=1}^N \overline{\mathbf{B}}_{i} (k,k') e^{-j2\pi(i-1)\frac{n-n'}{N}}+ V_{k,n}^{DD}.
	\end{array}\label{Y_k_n_exact}
\end{equation}
Let us now define the equivalent channel response in the Delay-Doppler domain as
\begin{equation}
	B^{DD}_{k,k',n,n'} =\sum_{i=1}^N \overline{\mathbf{B}}_{i} (k,k') e^{-j2\pi(i-1)\frac{n-n'}{N}}.
	\label{B_DD_def}
\end{equation}
Combining Eqs. \eqref{Y_k_n_exact} and \eqref{B_DD_def}, we obtain
\begin{equation}
	Y_{k,n}^{DD}= \frac{\sqrt{P_T}}{N} \ds \sum_{k'=0}^{K-1} \ds \sum_{n'=0}^{N-1} X_{k',n'}^{DD} B^{DD}_{k,k',n,n'} + V_{k,n}^{DD}
	\label{Y_k_n_2}
\end{equation}
Given \eqref{Y_k_n_2}, and denoting by $\mathbf{y}$ and $\mathbf{x}$  the vectorized versions of $\mathbf{Y}^{DD}$ and $\mathbf{X}^{DD}$, it can be finally shown that for the considered OTFS modulation with oversampled OTFS the following linear relationship holds:
\begin{equation}
	\mathbf{y}=\bm{\Psi} \mathbf{x}+\mathbf{v}
	\label{OTFS_I_O}
\end{equation}
where  $\mathbf{v}$ contains the AWGN contribution, and $\bm{\Psi}$ is a matrix whose entries are written as:
\begin{equation}
	\left[\bm{\Psi}\right](nK+k+1, n'K+k'+1)=\frac{\sqrt{P_T}}{N} B^{DD}_{k,k',n,n'},
	\label{Psi_matrix}
\end{equation}
with $k,k'=0,\ldots K-1$ and $n,n'= 0 ,\ldots, N-1$.

\section{The DR-UFMC modulation}
\begin{figure*}
	\begin{center}
		\includegraphics[scale=0.3]{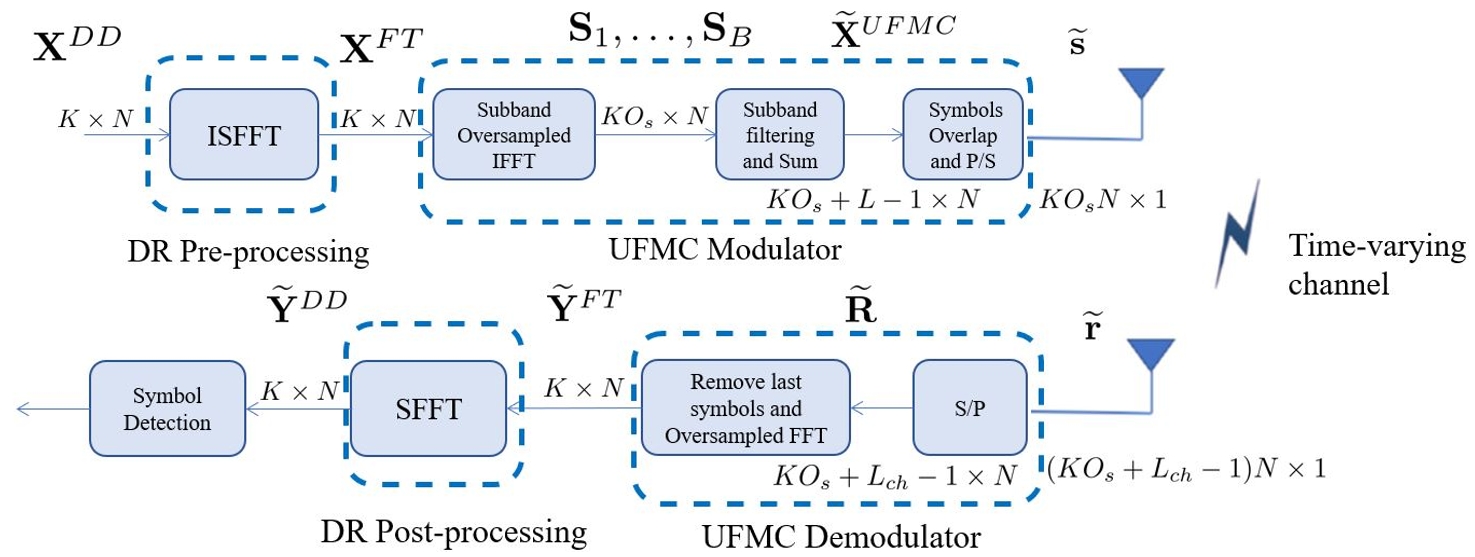}
	\end{center}
	\caption{DR-UFMC transceiver scheme consists of an UFMC transceiver with Doppler Resilient (DR) pre- and post-processing.}
	\label{Fig:DR_UFMC_v1}\end{figure*}

The transceiver block scheme of the newly proposed DR-UFMC modulation is reported in 
Fig. \ref{Fig:DR_UFMC_v1}. As it can be seen, an {UFMC transmitter is considered  in place of the OFDM transmitter} after the ISFFT operation. 
Again, we assume that the data symbols are arranged in the $(K \times N)$-dimensional matrix $\mathbf{X}^{DD}$ and that they go through the ISFFT transformation leading to  the matrix $\mathbf{X}^{FT}$. 
With regard to the UFMC modulator, we assume that the $K$ subcarriers are split in $B$ subbands of $D$ subcarriers each (thus implying that $K=BD$). 
Each subband is filtered using a passband FIR filter. A customary choice is to resort to Dolph-Chebyshev discrete-time window, that permits controlling the side-lobes' level with respect to the peak of the main lobe. 
The matrix $\mathbf{X}^{FT}$ is thus firstly processed by an oversampled IFFT on each subband of $D$ subcarriers. We  thus obtain $B$ matrices as follows
\begin{equation}
	\mathbf{S}_{i}=\overline{\mathbf{W}}_{K,O_s}^H \mathbf{P}_i  \mathbf{X}^{FT}
\end{equation}
with $i=0,\ldots, B-1$ and the selection matrices  $\mathbf{P}_i, \; i=0, \ldots, B-1$ are defined as
\begin{equation}
	\mathbf{P}_i =\mbox{diag}\left([\; \underbrace{0 \ldots 0}_{iD} \underbrace{1 \ldots 1}_D \underbrace{0 \ldots 0}_{K-(i+1)D}]\right) \; , \quad i=0, \ldots, B-1 \; .
\end{equation}
Then each subband is individually filtered via the Toeplitz matrices $\mathbf{G}_i$, with $i=0,\ldots, B-1$.
Precisely, letting $\mathbf{g}= [g_0, g_1, \ldots, g_{L-1}]^T$ be the $L$-dimensional vector representing the Dolph-Chebyshev prototype filter, and letting $F_i = \frac{D-1}{2} + iD - \frac{K}{2}$ denote the normalized frequency shift of the filter tuned to the $i$-th subband, $\mathbf{G}_i$ is a 
Toeplitz $[(KO_s+L-1) \times  KO_s]$-dimensional matrix describing the discrete convolution operation with the filter $\mathbf{g}_i$,  whose $\ell$-th element is defined as
$
	g_{i,\ell}=g_i e^{j 2 \pi \frac{F_i \ell}{KO_s}}$, for $i=0, \ldots, B-1$, and $\ell=0, \ldots, L-1$.
The vectors at the output of the subband filters are then added, leading to the $[(KO_s+L-1) \times  N]$-dimensional matrix $\widetilde{\mathbf{X}}^{UFMC}$
\begin{equation}
	\widetilde{\mathbf{X}}^{UFMC}=\ds\sum_{i=0}^{B-1}\mathbf{G}_i \overline{\mathbf{W}}_{K,O_s}^H \mathbf{P}_i  \mathbf{X}^{FT} = \mathbf{P}_{UFMC} \mathbf{X}^{FT},
	\label{X_tilde_UFMC}
\end{equation}
where we have defined $
	\mathbf{P}_{UFMC}=\sum_{i=0}^{B-1}\mathbf{G}_i \overline{\mathbf{W}}_{K,O_s}^H \mathbf{P}_i $.
To obtain a $(K O_s \times N)$-dimensional matrix, $\mathbf{X}^{UFMC}$ say, we rely on the concept of continuous packet transmission in UFMC as introduced in \cite{buzzi2019mimo}: no guard time among consecutive packets is thus assumed in order to increase the modulation spectral efficiency. The 
 $(k,n)-$th entry of the resulting matrix $\mathbf{X}^{UFMC}$ is reported in  Eq.  \eqref{X_tilde_overlap}, shown at the top of the next page, $\forall k=1,\ldots KO_s$ and $n=1,\ldots, N$.
\begin{figure*}[t]
\begin{equation}
	\mathbf{X}^{UFMC}(k,n)=\left \lbrace
	\begin{array}{lll}
		\widetilde{\mathbf{X}}^{UFMC}(k,n) \; \;  \text{if} \, (n=1 \, \text{and} \, k\leq K O_s) \, \text{or} \, (n>1 \, \text{and} \, L \leq k\leq KO_s) 
		\\
		\widetilde{\mathbf{X}}^{UFMC}(k,n) + \widetilde{\mathbf{X}}^{UFMC}(k+K,n-1) \; \; \text{if} (n>1 \, \text{and} \,  k \leq L-1)
	\end{array} \right. 
	\label{X_tilde_overlap}
\end{equation} \begin{center}\rule{10cm}{0.3mm} \end{center}
\end{figure*}
Note that in Eq. \eqref{X_tilde_overlap} we are removing the last $L-1$ symbols of the last column of $\widetilde{\mathbf{X}}^{UFMC}$  in order to transmit a 1-D signal with length $KO_sN$ as it happens with the OTFS modulation.
To this end, we will use later the matrix $\overline{\mathbf{I}}_{L-1}=\left[\mathbf{I}_{KO_sN} , \mathbf{0}_{KO_sN \times L-1}\right]$. 
The 1-D signal  $\widetilde{\mathbf{s}}$ to be transmitted is thus:
\begin{equation}
	\widetilde{\mathbf{s}}=\text{vec}\left( \mathbf{X}^{UFMC}\right) \; .
\end{equation} 
Let us now define 
$\widetilde{\mathbf{U}}_{UFMC}$ as a $[(KO_sN+L-1) \times KO_sN]$-dimensional matrix with the following $[(KO_s+L-1)\times KO_s]$-dimensional non-zero blocks
\begin{equation}
	\begin{array}{ll}
		\widetilde{\mathbf{U}}_{UFMC}((n-1)KO_s+1:nKO_s+L-1, \\ (n-1)KO_s+1:nKO_s)=\mathbf{P}_{UFMC},
		\label{U_tilde_definition}
	\end{array}
\end{equation}
with $n=0,\ldots, N-1$. 
Define also $\mathbf{U}_{UFMC}=\overline{\mathbf{I}}_{L-1} \widetilde{\mathbf{U}}_{UFMC}$.
Using Eqs. \eqref{X_tilde_UFMC} and \eqref{X_ISFFT}, it can be shown that $\widetilde{\mathbf{s}}$ can be written as 
\begin{equation}
	\widetilde{\mathbf{s}}= \mathbf{U}_{UFMC} \text{vec}\left( \mathbf{X}^{FT} \right) = \mathbf{U}_{UFMC} \text{vec}\left(\mathbf{F}_K \mathbf{X}^{DD} \mathbf{F}_N^H\right) 
	\label{s_vec_DR_UFMC}
\end{equation}
Exploiting the properties of the vec$(\cdot)$ operator Eq. \eqref{s_vec_DR_UFMC} can be written as
\begin{equation}
	\widetilde{\mathbf{s}}= \mathbf{U}_{UFMC} \left(\mathbf{F}_N^* \otimes \mathbf{F}_K\right) \mathbf{x}, 
	\label{s_vec_DR_UFMC_v1_2}
\end{equation}
where $\mathbf{x} =\text{vec}\left(\mathbf{X}^{DD}\right)$ and $\otimes$ denotes the Kronecker product.
The signal $\widetilde{\mathbf{s}}$ is thus transmitted on the time-varying channel already described in the previous section.
The $(KO_s+L_{ch}-1) N$-th dimensional received signal $\widetilde{\mathbf{r}}$ can be written as
\begin{equation}
	\widetilde{\mathbf{r}}= \sqrt{P_T} \widetilde{\mathbf{M}}_{h,N}\widetilde{\mathbf{s}}+\widetilde{\mathbf{w}},
\end{equation}
where $\widetilde{\mathbf{w}}$ is the AWGN contribution, and  $\widetilde{\mathbf{M}}_{h,N}=\text{blckdiag}\left(\widetilde{\mathbf{M}}_{h}^{(1)},\ldots, \widetilde{\mathbf{M}}_{h}^{(N)} \right)$.
The entries of the $\left[(KO_s+L_{ch}-1) \times (KO_s)\right]$-dimensional matrix $\widetilde{\mathbf{M}}_{h}^{(i)}$ can be expressed as in Eq. \eqref{M_h_i_LTV_channel}.
At the receiver, the  matrix $\widehat{\mathbf{R}} $ of size $KO_s+L_{ch}-1\times N$ is formed, i.e., $\widehat{\mathbf{R}} = \text{invec}\{\widetilde{\mathbf{r}}\}$,
where each column vector of $\widehat{\mathbf{R}}$ can be interpreted as a received UFMC symbol:
\begin{equation}
	\widehat{\mathbf{R}}(:,i) = \sqrt{P_T} \widetilde{\mathbf{M}}_{h}^{(i)}  \mathbf{X}^{UFMC}(:,i) + \widehat{\mathbf{W}}(:,i)
\end{equation}
with $i=1,\ldots,N$ and $\widehat{\mathbf{W}} = \text{invec}\{\widetilde{\mathbf{w}}\}$.
Using the classical UFMC processing \cite{buzzi2019mimo}, the last $L_{ch}-1$ symbols are removed from the received signal  $\widehat{\mathbf{R}}$. Defining
	$\overline{\mathbf{R}}_{L_{ch}-1}=\left[ \mathbf{I}_{KO_s} \; \, \mathbf{0}_{KO_s,L_{ch}-1}\right]$,
the cancellation of such symbols results in $\widetilde{\mathbf{R}}=\overline{\mathbf{R}}_{L_{ch}-1}\widehat{\mathbf{R}}$.
It can be shown that the $n'$-th column of $\widetilde{\mathbf{R}}$ is expressed as
\begin{equation}
	\widetilde{\mathbf{R}}(:,n')=  \sqrt{P_T} \widetilde{\mathbf{H}}_{n'} \mathbf{X}^{UFMC}(:,n') + \widetilde{\mathbf{V}}(:,n')
	\label{R_n_DR_UFMC}
\end{equation}
with $\widetilde{\mathbf{V}}=\overline{\mathbf{R}}_{L_{ch}-1}\widehat{\mathbf{W}}$ and $\widetilde{\mathbf{H}}_{n'}=\overline{\mathbf{R}}_{L_{ch}-1}\widetilde{\mathbf{M}}_{h}^{(n')}$.
Applying the oversampled FFT operation on each UFMC symbol, we obtain the received $K \times N$- dimensional matrix $\widetilde{\mathbf{Y}}^{FT}$ in the frequency-time domain as
$	\widetilde{\mathbf{Y}}^{FT} = \overline{\mathbf{W}}_{K,O_s} \widetilde{\mathbf{R}}$.
Next, we apply the SFFT, which results in
\begin{equation}
	\begin{array}{lll}
	\widetilde{\mathbf{Y}}^{DD}= & \mathbf{F}_K^H \widetilde{\mathbf{Y}}^{FT} \mathbf{F}_N= \mathbf{F}_K^H \overline{\mathbf{W}}_{K,O_s} \widetilde{\mathbf{R}}\mathbf{F}_N= \\ & \ds \sum_{n'=0}^{N-1} \mathbf{F}_K^H \overline{\mathbf{W}}_{K,O_s} \widetilde{\mathbf{R}}(:,n') \mathbf{F}_N(:,n')^T
	\end{array}
    \label{Y_DD_DR_UFMC}
\end{equation}
By substituting \eqref{R_n_DR_UFMC} in \eqref{Y_DD_DR_UFMC}, we obtain
\begin{equation}
	\begin{array}{lll}
	\widetilde{\mathbf{Y}}^{DD}=\\ \sqrt{P_T}\ds \sum_{n'=0}^{N-1} \mathbf{F}_K^H \overline{\mathbf{W}}_{K,O_s} \widetilde{\mathbf{H}}_{n'} \mathbf{X}^{UFMC}(:,n')\mathbf{F}_N(:,n')^T  + \widetilde{\mathbf{V}}_{DD}
	\end{array}\label{YY_DD_DR_UFMC_2}
\end{equation}
with $\widetilde{\mathbf{V}}_{DD}=\widetilde{\mathbf{V}}\mathbf{F}_N$. Upon defining matrix the $(KO_s\times N)$-dimensional $\mathbf{A}_{n'}=\mathbf{X}^{UFMC}(:,n')\mathbf{F}_N(:,n')^T$, we can show that its entries can be written as
\begin{equation}
	\mathbf{A}_{n'}(\ell,m)=\ds \frac{1}{\sqrt{N}} \mathbf{X}^{UFMC}(\ell,n') e^{-j 2 \pi \frac{nm}{N}},
	\label{A_n_def}
\end{equation}
$n'=0,\ldots, N-1$, $ \ell=0,\ldots, KO_s-1$ and $m=0,\ldots, N-1$.
Upon some algebraic manipulations, we obtain that the $(k,n)$-th entry of the received matrix $\widetilde{\mathbf{Y}}^{DD}$ can be obtained as 
\begin{equation}
	\begin{array}{lll}
		\widetilde{\mathbf{Y}}^{DD}(k,n)=\\ \sqrt{P_T} \ds \sum_{n'=0}^{N-1} \ds \sum_{k'=0}^{KO_s-1} \mathbf{F}_K^H \overline{\mathbf{W}}_{K,O_s} \widetilde{\mathbf{H}}_{n'} (k,k') \mathbf{A}_{n'}(k',n) + \mathbf{V}_{DD}(k,n)\\ =\ds\frac{\sqrt{P_T}}{\sqrt{N}} \ds \sum_{n'=0}^{N-1} \ds \sum_{k'=0}^{KO_s-1} \widetilde{\mathbf{B}}_{n'}(k,k') \mathbf{X}^{UFMC}(k',n')e^{-j 2 \pi \frac{nn'}{N}}\\  + \widetilde{\mathbf{V}}_{DD}(k,n)
		\label{YY_DD_DR_UFMC_3}
	\end{array}
\end{equation}
with $k=0,\ldots, K-1$, $n=0,\ldots ,N-1$ and $
	\widetilde{\mathbf{B}}_{n'}=\mathbf{F}_K^H \overline{\mathbf{W}}_{K,O_s} \widetilde{\mathbf{H}}_{n'}$.
Manipulating \eqref{YY_DD_DR_UFMC_3} and using Eq. \eqref{s_vec_DR_UFMC_v1_2}, we obtain for the DR-UFMC modulation the following  input/output relationship between the received data and the transmitted data symbols:
\begin{equation}
	\widetilde{\mathbf{y}}=\widetilde{\bm{\Psi}} \mathbf{x} +\widetilde{\mathbf{v}}
	\label{DR_UFMC_v1_I_O}
\end{equation}
where $\widetilde{\mathbf{y}}$ and $\mathbf{x}$ are the vectorized versions of $\widetilde{\mathbf{Y}}^{DD}$ and $\mathbf{X}^{DD}$, respectively, and $\widetilde{\mathbf{v}}$ contains the AWGN contribution. In \eqref{DR_UFMC_v1_I_O},  $\widetilde{\bm{\Psi}}$ is defined as
\begin{equation}
	\widetilde{\bm{\Psi}}=\bm{\Psi}_{UFMC} \mathbf{U}_{UFMC} \left(\mathbf{F}_N^* \otimes \mathbf{F}_K\right),
	\label{Psi_v1_def}
\end{equation}
where we used Eq. \eqref{s_vec_DR_UFMC_v1_2} and the entries of $\bm{\Psi}_{UFMC}$ are written as:
\begin{equation}
	\begin{array}{lll}
	\left[\bm{\Psi}_{UFMC}\right](nK+k+1, n'KO_s+k'+1)= \\ \qquad
	 \ds\frac{\sqrt{P_T}}{\sqrt{N}} \widetilde{\mathbf{B}}_{n'}(k,k') e^{-j2\pi \frac{nn'}{N}},
	\end{array}
\label{Psi_UFMC_v1_def}
\end{equation}
with $k=0,\ldots K-1$, $k'=0,\ldots KO_s-1$ and $n,n'=0, \ldots, N-1$.

\section{Performance measures and results}
Table \ref{table:parameters_2} shows the considered simulation setup. We use the Extended Vehicular A (EVA) channel model \cite{hong2022delay}. Let $\nu_{\rm max}= \frac{f v_{\rm max}}{c}$ be the maximum Doppler shift, with $f$ the frequency in Hz, $v_{\rm max}$ the user's speed in m/s and $c$ the speed of light in m/s. We assume that a single Doppler shift is associated with the $p$-th path and follows the classic Jakes spectrum, i.e., $\nu_p=\nu_{\rm max} \cos\left( \theta_p\right)$ where $\theta_p$ is uniformly distributed over $[-\pi, \pi]$.

\begin{table}[t]
	\centering
	\caption{Simulation parameters}
	\label{table:parameters_2}
	\def\arraystretch{1.2}
	\begin{tabulary}{\columnwidth}{ |p{1.7cm}|p{4cm}|p{1.5cm}| }
		\hline
		\textbf{Name} 				& \textbf{Meaning} & \textbf{Value}\\ \hline
		$f$	&  Carrier frequency & 28 GHz\\ \hline
		$\Delta_f$	&  subcarrier spacing & 120 KHz\\ \hline
		$K$ 				& number of subcarriers  & 128\\ \hline
		$N$ 				& number of symbols  & 16\\ \hline
		$T_{CP}$ 				& duration of the cyclyc prefix in OTFS & 0.586 $\mu$s\\ \hline
		$W=K \Delta_f$ 				& system bandwidth & 15.36 MHz\\ \hline
		$T=1/\Delta_f$ 				& symbol interval & 8.33 $\mu$s\\ \hline
		$O_s$ 				& oversampling factor & 10 \\ \hline
		$L$ 				& length of the Dolph-Chebyshev FIR filter in DR-UFMC & 60 \\ \hline		
		$A_{dB}$ 				& OOB attenuation of the Dolph-Chebyshev FIR filter in DR-UFMC & 100 \\ \hline
		$D$ 				& number of subcarrier in each subband in DR-UFMC & 16 \\ \hline
		$\delta_{OOB}$ 			& OOB threshold & -30 dB \\ \hline
	\end{tabulary}
\end{table}

\begin{figure}
	\centering
	\includegraphics[scale=0.32]{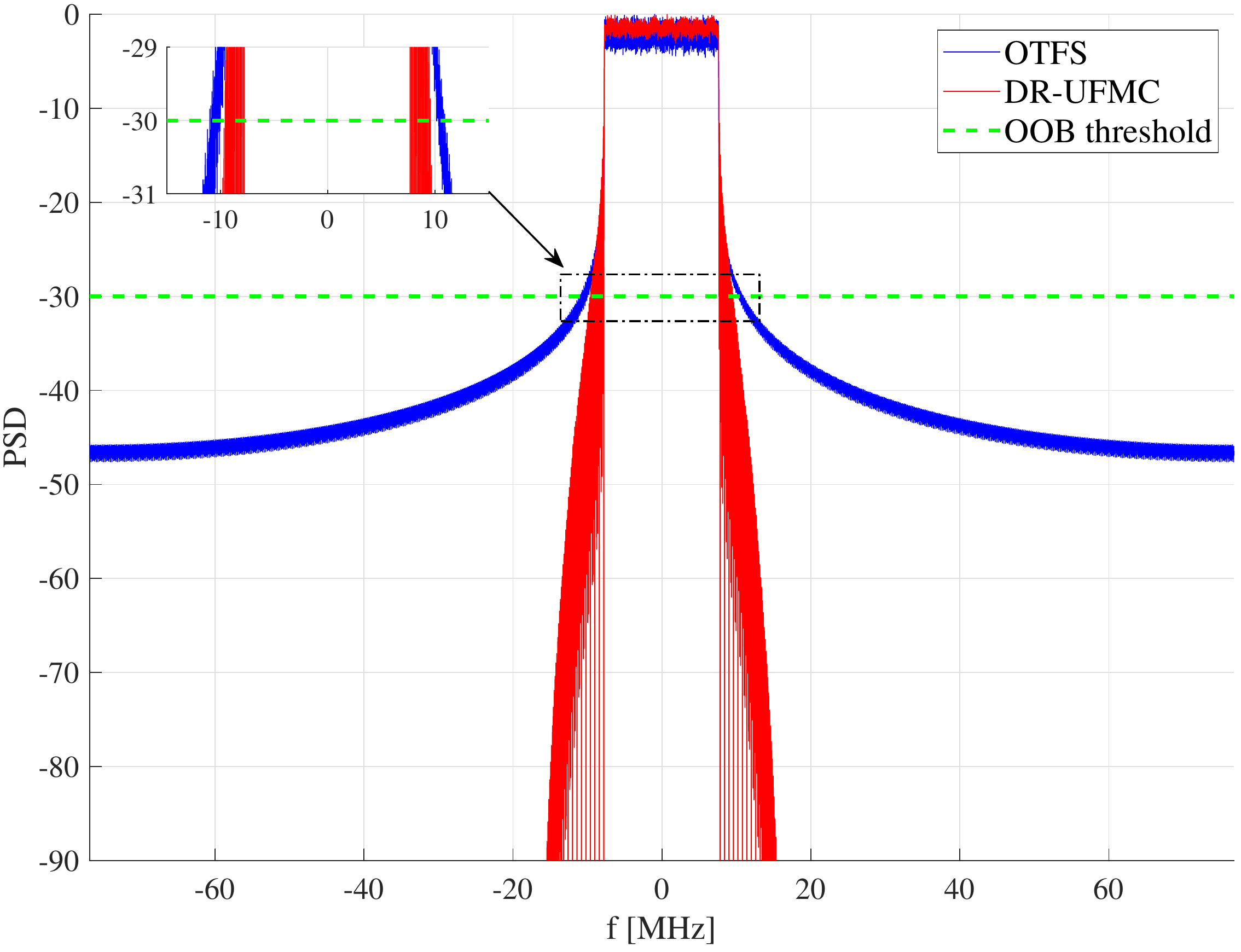}
	\caption{PSD of DR-UFMC and OTFS for the parameters of  Table \ref{table:parameters_2}.}
	\label{Fig:PSD_comparison}
\end{figure}

First of all, we have to investigate on the power-spectral-density of the OTFS and DR-UFMC signals in order to carry out a performance comparison with the same level of OOB emissions for both modulations. 
Fig. \ref{Fig:PSD_comparison} shows the PSD of OTFS and DR-UFMC transmitted signal. 
It is clearly seen that the spectral occupation of DR-UFMC is considerably lower than OTFS:
the number of subcarriers at the edge, say $2N_G$, to be nulled in order to guarantee that the emissions out of the nominal bandwidth $[-K\Delta_f/2, K\Delta_f/2]$ are below the OOB threshold $\delta_{OOB}=-30$ dB is 60 for the OTFS modulation (and also for OFDM) and 36 for the DR-UFMC modulation. 

Next, we provide the definition of the Spectral Efficiency (SE) associated to the $k$-th subcarrier in the $n$-th symbol, i.e.:
\begin{equation}
	\text{SE}_{k,n}= \xi \log_2 \left(1+ \text{SINR}_{k,n}\right)
\end{equation}
where $\text{SINR}_{k,n}$ is the corresponding SINR (set to zero if $k$ is one of the guard subcarriers to be nulled due to the OOB emissions constraint) and $\xi$ is the efficiency factor of the modulation scheme.  For OTFS, we have 
$\xi=  \frac{T}{T+T_{CP}}$, where the symbols $T$ and $T_{CP}$ are defined in Table \ref{table:parameters_2}. For DR-UFMC, instead, $\xi= 1$ since, according to the scheme in \cite{buzzi2019mimo}, no CP is used. 
The {spectral efficiency averaged over all the subcarriers and symbols} can be thus defined as
$\overline{\text{SE}}= \frac{\xi}{KN} \sum_{k=1}^K  \sum_{n=1}^N \log_2 \left(1+ \text{SINR}_{k,n}\right)$.
Finally, we provide the SINRs expressions. The input-output linear relationships \eqref{OTFS_I_O} and \eqref{DR_UFMC_v1_I_O}, can be written as
\begin{equation}
	\overline{\mathbf{y}}=\mathbf{C}\mathbf{x}+ \mathbf{w}=\ds \sum_{i=1}^{NK} {\mathbf{C}_i x_i} + \mathbf{w},
	\label{I_O_generic}
\end{equation}
where, for OTFS $\overline{\mathbf{y}}=\mathbf{y}$, $\mathbf{C}=\bm{\Psi}$ and $\mathbf{w}=\mathbf{v}$, while for DR-UFMC  $\overline{\mathbf{y}}=\widetilde{\mathbf{y}}$, $\mathbf{C}=\widetilde{\bm{\Psi}}$ and $\mathbf{w}=\widetilde{\mathbf{v}}$.
In Eq. \eqref{I_O_generic} $\mathbf{C}_i$ denotes the $i$-th column of the matrix $\mathbf{C}$.
Resorting to linear MMSE detection, the estimate of the symbol transmitted on the $k$-th subcarrier in the $n$-th symbol can be written as
\begin{equation}
	\begin{array}{lll}
	\hat{x}_{(n-1)K+k}= \underbrace{\mathbf{C}_{(n-1)K+k}^H  \left( {\mathbf{C} \mathbf{C}^H + \sigma^2_w \mathbf{I}_{KN}} \right)^{-1}}_{
	\triangleq \mathbf{d}_{(n-1)K+k}^H} \overline{\mathbf{y}}\, ,
\end{array}\end{equation}
with $\sigma^2_w \mathbf{I}_{KN}$ the covariance matrix of noise.
The SINR expression on the $k$-th subcarrier in the $n$-th symbol can be thus written as follows
\begin{equation}
	\text{SINR}_{k,n}=  \frac{\left|\mathbf{d}_{(n-1)K+k}^H \mathbf{C}_{(n-1)K+k}\right|^2}{ \!\!\!\sum_{\substack{j=1 \\ j \neq (n-1)K+k}}^{NK}\left|\mathbf{d}_{(n-1)K+k}^H \mathbf{C}_j\right|^2\!\! \!\! + \!\sigma^2_w \|\mathbf{d}_{(n-1)K+k}\|^2 }
	\label{SINR_MMSE}
\end{equation}
Given \eqref{SINR_MMSE}, the {net SINR per subcarrier and symbol} can be defined as the average SINR across the non-null subcarriers:
\begin{equation}
	\text{SINR}_{N_G}=\frac{1}{(K-2N_G)N} \sum_{k=N_G+1}^{K-N_G}  \sum_{n=1}^N\text{SINR}_{k,n}.
	\label{Net_SINR}
\end{equation}

\begin{figure}
	\centering
	\includegraphics[scale=0.35]{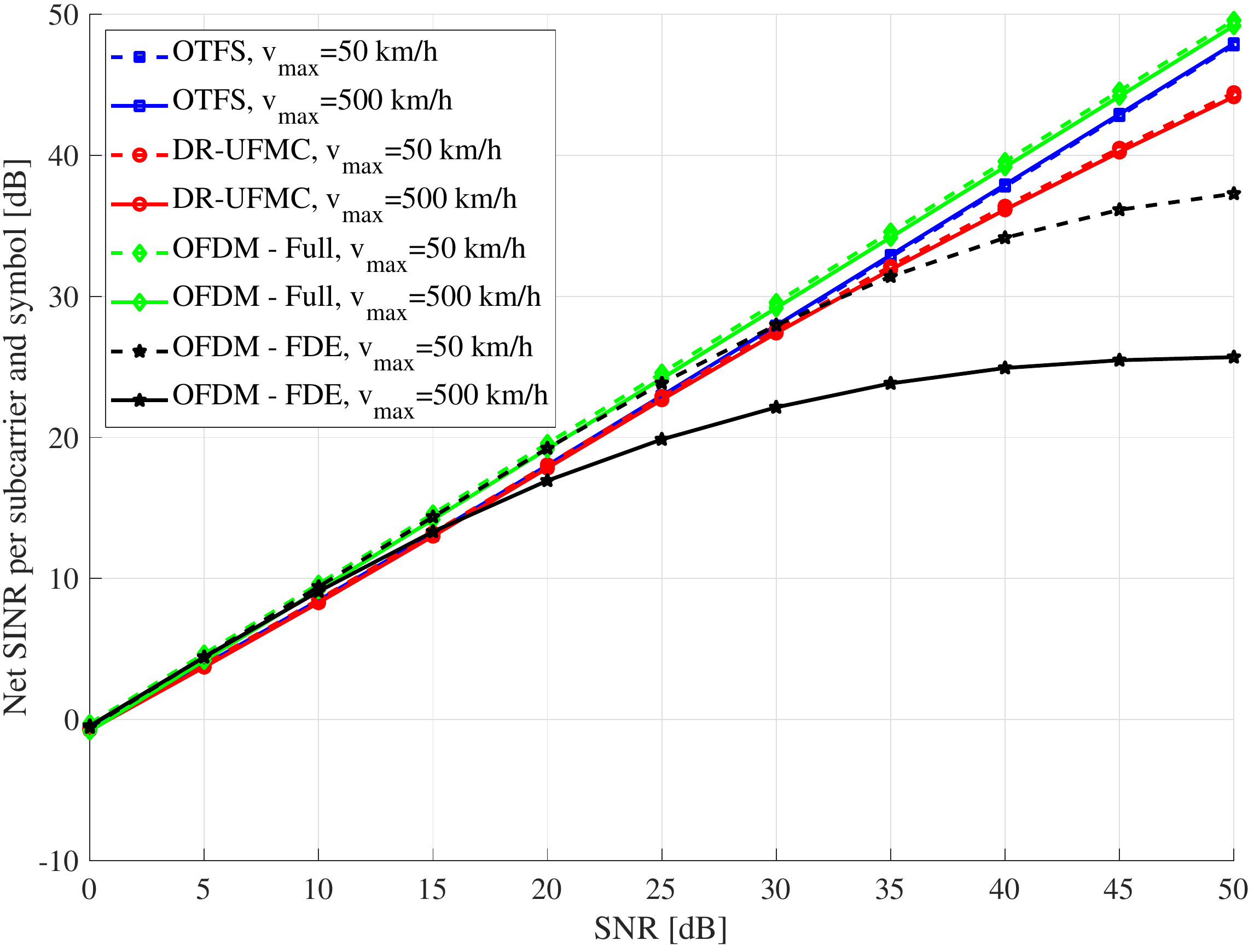}
	\caption{Net SINR per subcarrier and symbol versus SNR, for two different values of radial velocity.}
	\label{Fig:SINR_comparison}
\end{figure}

\begin{figure}
	\centering
	\includegraphics[scale=0.35]{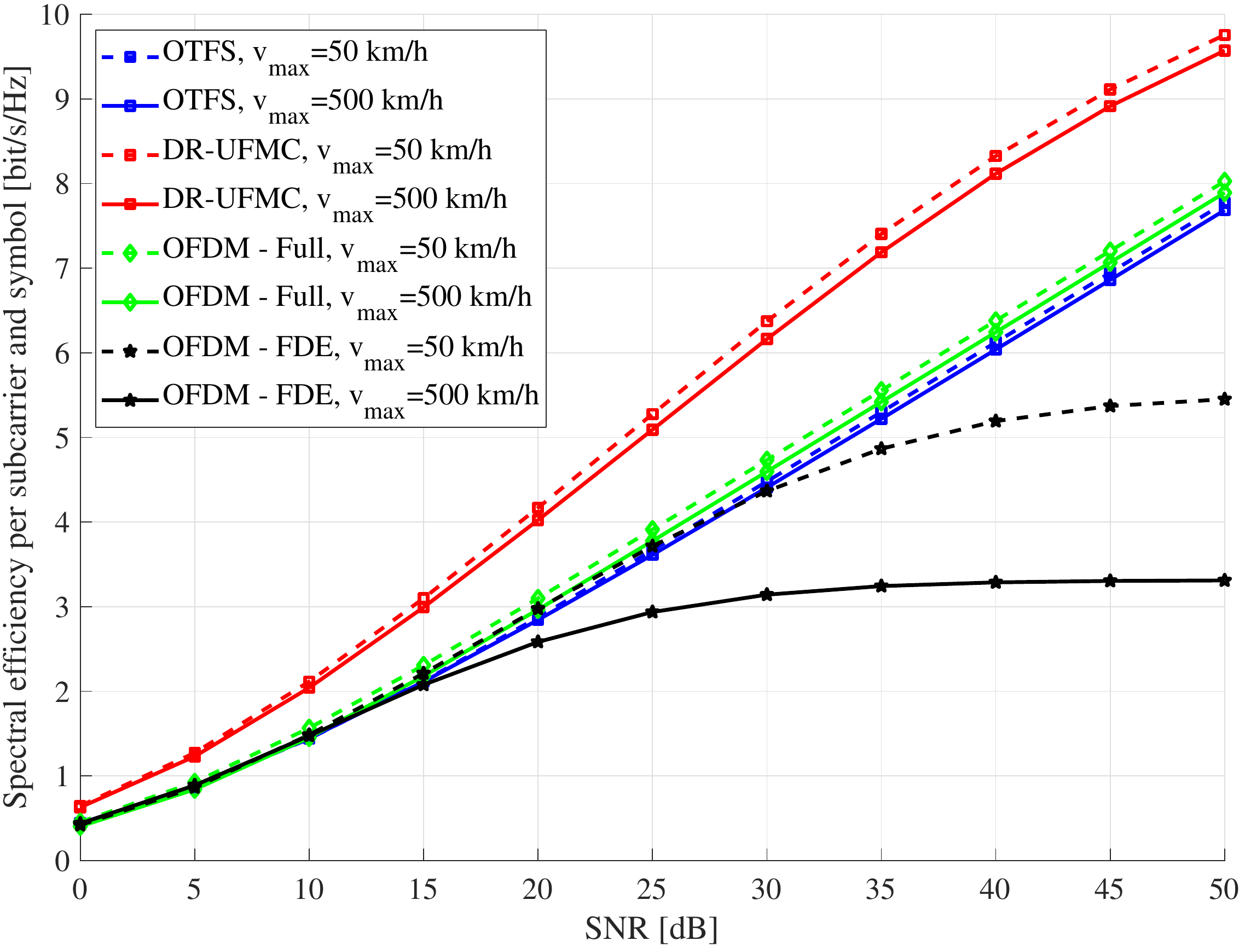}
	\caption{Average Spectral Efficiency per subcarrier and symbol  versus SNR, for two different values of radial velocity.}
	\label{Fig:SE_comparison}
\end{figure}

Figs. \ref{Fig:SINR_comparison} and \ref{Fig:SE_comparison} show the performance in terms of net SINR and of average spectral efficiency per subcarrier and symbol versus the SNR, with the aim of comparing DR-UFMC, OTFS,  OFDM with full multicarrier multisymbol processing\footnote{This descends from the OTFS derivation in the special case in which the ISFFT and the SFFT blocks are removed.}, and OFDM with one-tap FDE. Two different values of the maximum speed of the user $v_{max}$ are chosen, namely 50 km/h and 500 km/h. 
Inspecting the figures, the following conclusions can be drawn: (a) in terms of net SINR, OFDM with full processing, OTFS and DR-UFMC achieve approximately the same performance, while OFDM with one-tap FDE exhibits an heavy performance degradation, especially for large values of $v_{max}$; (b) OTFS, DR-UFMC and OFDM with full processing are almost insensitive to the value of the maximum speed; (c) in terms of SE, instead, the net superiority of DR-UFMC with respect to the other competing alternatives is clearly seen; again, OFDM with one-tap FDE achieves the worst performance. 
\section{Conclusions}
This paper has introduced a novel modulation format, named DR-UFMC,  which is to be intended as an evolution of OTFS. This modulation has been shown to exhibit both robustness to Doppler shifts and good properties of spectral containment, so it is to be considered a better alternative to OFDM than OTFS. Current research is focused on the derivation of suitable channel estimation schemes and data detectors for the newly introduced modulation scheme.

\section*{Acknowledgment}
This work was supported by Huawei Technologies Duesseldorf GmbH through cooperation agreement TC20220209479.

\begin{small}
\bibliography{BEMOD_refs}
\bibliographystyle{IEEEtran}
\end{small}

\end{document}